\documentclass[english]{ieee}
\usepackage[T1]{fontenc}
\usepackage[latin1]{inputenc}
\usepackage{graphicx}
\usepackage{setspace}

\makeatletter

\providecommand{\tabularnewline}{\\}

\usepackage{babel}
\makeatother
\begin{document}

\title{Performance Characterisation of Intra-Cluster Collective Communications}

\author{Luiz Angelo Barchet-Estefanel\thanks{Supported by grant BEX 1364/00-6 from CAPES - Brazil},
Grégory Mounié\\
 ID - IMAG Laboratory, APACHE Project\thanks{This project is supported by CNRS, INPG, INRIA and UJF}\\
51, Avenue Jean Kuntzmann, F38330 Montbonnot St. Martin, France\\
\{Luiz-Angelo.Estefanel,Gregory.Mounie\}@imag.fr}

\maketitle
\begin{abstract}
Although recent works try to improve collective communication in grid
systems by separating intra and inter-cluster communication, the optimisation
of communications focus only on inter-cluster
communications. We believe, instead, that the overall performance
of the application may be improved if intra-cluster collective communications
performance is known in advance. Hence, it is important to have an accurate
model of the intra-cluster collective communications, which provides
the necessary evidences to tune and to predict their performance correctly. 
In this paper we present our experience on
modelling such communication strategies. We describe and compare different
implementation strategies with their communication models, evaluating
the models' accuracy and describing the practical challenges that
can be found when modelling collective communications. 
\end{abstract}
\vspace{-0.3cm}
\section{\label{sec:Introduction}Introduction}

The optimisation of collective communications in grids is a complex
task because the inherent heterogeneity of the network limits the
use of general solutions. To reduce the complexity cost, most systems
consider grids as interconnected \emph{islands of homogeneous clusters}.
Although there are no restrictions on the number of layer that connect
those ``islands'', as successfully demonstrated by \cite{key-31},
most systems only optimise communications at the inter-cluster level, because
wide-area networks are slower than LANs. 
Some examples of this ``two-layered''
approach include ECO \cite{key-5}, MagPIe \cite{key-6}\cite{key-9},
that apply this concept for wide-area networks, and even LAM-MPI 7
\cite{key-32}, that consider SMP machines as islands of fast communication. 

We believe that while inter-cluster optimisation is necessary to achieve
good performances in grid-like environments, its optimisation should not be
disconnected from the intra-cluster level.
Actually, the modelling and optimisation of intra-cluster
communication is specially important when the clusters are
structured in multiple layers. In this situation, the grid-aware tools
must deal with both communication and topology mapping, and a \emph{priori} 
knowledge on the intra-clusters communication may lead to more important
reductions of the overall execution time than a simple minimisation of the 
wide-area communications. 

Hence, in this paper we investigate how performance models can be used to 
characterise the communication patterns of the collective communications.
These models can be used both to predict the performance of these operations 
and to decide which implementation technique is the better adapted for a 
specific set of parameters (number of processes,
message size, network performance, etc.). 

Consequently, to model collective communications we need a good performance
model. There are several performance models for message-passing parallel
programs, some of them widely known like BSP \cite{key-39}
or LogP \cite{key-3}. Although these two models are equivalent in
most circumstances \cite{key-42}, LogP is slightly more general than
BSP, as it does not requires a global barrier to separate communication
and computation phases, and because it adds the notion of finite network
capacity that can only support a certain number of messages in transit
at once. As consequence, we choose to use, in this paper, the \emph{parameterised
LogP} model \cite{key-9}. pLogP is an extension of the LogP model
that can accurately handle both small messages and large messages
with a low complexity. Due to its simplicity, this model allows a
fast prototyping of the communication performance, even though it
has difficulties to represent contention situations. Nevertheless,
our pLogP models were able to predict with enough accuracy the system
performance in most cases presented in this paper, allowing the selection
of the most adapted implementation technique to a specific network
environment.

To illustrate our approach, we present three examples, the Broadcast,
Scatter and All-to-All operations, which respectively represent the
``one-to-many'', ``personalised one-to-many'' and ``many-to-many''
collective communications. While conceptually simple, Broadcast and
Scatter operations have communication patterns that can be found in
many other operations, like Barriers, Reduces and Gathers. The All-to-All
operation, instead, has a complex communication pattern, but is one
of the most important communication patterns for scientific applications.
Additionally, an All-to-All operation is subjected to important problems
with communication contention, representing a real challenge to performance
modelling. 

The rest of this paper is organised as follows: Section \ref{sec:System_Model}
presents the definitions and the test environment we will consider
along this paper. Sections \ref{sec:Broadcast}, \ref{sec:Scatter}
and \ref{sec:Alltoall} present, respectively the communication models
we developed for both Broadcast, Gather and All-to-All, while comparing
the predictions from those models with experimental results.
Finally, Section \ref{sec:Conclusions}
presents our conclusions, as well as the future directions of the
research.

\section{\label{sec:System_Model}System Model and Definitions}

In this paper we model collective communications using the \emph{parameterised
LogP} model, or simply pLogP \cite{key-9}. As pLogP parameters
depend on the message size, it can be accurate
when dealing with both small and large messages. Further, the paper
that describes pLogP presents several communication models for grid-aware
collective communications, which served as guide to many of our own
communication models. 

Therefore, all along this paper we shall use the same terminology
from pLogP's definition, such as \emph{g(m)} for the gap of a message
of size \emph{m}, \emph{L} as the communication latency between two
nodes, and \emph{P} as the number of nodes. In the case of message
segmentation, the segment size \emph{s} of the message \emph{m} is
a multiple of the size of the basic datatype to be transmitted, and
it splits the initial message \emph{m} into \emph{k} segments. Thus,
\emph{g(s)} represents the gap of a segment with size \emph{s}.

\begin{figure}
\begin{center}\includegraphics[%
  width=1.0\columnwidth]{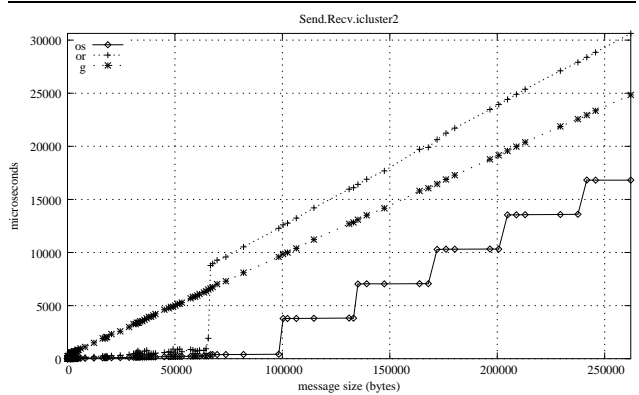}\end{center}
  \vspace{-0.3cm}

\caption{\label{fig:pLogP-parameters}pLogP parameters for the icluster-2 network}
\end{figure}

The pLogP parameters used to feed our models were obtained with the
MPI LogP Benchmark tool \cite{key-7} using LAM-MPI 7.0.4 \cite{key-32},
and are presented in Figure \ref{fig:pLogP-parameters}.

The experiments to obtain pLogP parameters, as well as the practical
experiments, were conducted on the \textbf{ID/HP icluster-2} from
the ID laboratory Cluster Computing Centre%
\footnote{http://www-id.imag.fr/Grappes/%
}. This cluster contains 100 Itanium-2 (IA-64) machines (Dual processor,
900MHz, 3GB) interconnected by a switched Ethernet 100 Mbps network,
running Red Hat Linux Advanced Server 2.1AS with kernel 2.4.18smp.
The experiments consisted on 100 measures for each set of parameters (message 
size, number of processes), and the values presented here are the average of such measures.

\section{\label{sec:Broadcast}One-to-Many: Broadcast}

With Broadcast, a single process, called \emph{root,} sends the same
message of size \emph{m} to all other $(P-1)$ processes. Classical
implementations of the Broadcast operation rely on \emph{d-}ary trees
characterised by two parameters, \emph{d} and \emph{h}, where \emph{d}
is the maximum number of successors a node can have, and \emph{h}
is the height of the tree, the longest path from the root to any of
the tree leaves. While an optimal tree shape can be deduced from
the network parameters and from \emph{d, h $\in$}{[1...\emph{P}-1]}
for which $\sum_{i=o}^{h}d^{i}\geq P$ is true, most MPI implementations
usually rely on two fixed shapes, the Flat Tree, for small number
of nodes, and the Binomial Tree. 

Because most MPI implementations rely only on Flat and Binomial Broadcast,
some techniques were developed to improve its efficiency. This way,
it is usual to apply different strategies according to the message
size, as for example, the use of a \emph{rendezvous} message that
prepares the receiver to the incoming of a large message, or the use
of non-blocking primitives to overlap communication and computation.
Unfortunately, such techniques bring only minimal improvements to
the final performance, and their efficiency still depends mostly on
the network characteristics.

Another possibility, however, is to compose a Chain among the processes,
pipelining messages \cite{key-33}. This strategy benefits from 
the use of message segmentation, presenting many advantages as recent works 
indicate \cite{key-9}\cite{key-34}.
In a Segmented Chain Broadcast, 
the transmission of messages in segments allows a node to overlap
the transmission of segment \emph{k} and the reception of segment
\emph{k}+1, reducing the overall \emph{gap} time. 

However, the size of the segments should be carefully chosen according to the 
network environment. Indeed, too small messages pay more for their
headers than for their content, while too large messages do not explore
enough the network bandwidth. The search for the segment size \emph{s}
that minimises the communication time can be done using the communication
models presented on Table \ref{table:bcast_models} and the network
parameters. An efficient method consists in searching through all
values of \emph{s} such that $s=m/2^{i},i\in[0\ldots log_{2}m]$.
To refine the search, we can also apply some heuristics like local
hill-climbing, as proposed by Kielmann \emph{et al.} \cite{key-9}.

In our work we developed the communication models for some current
techniques and their ``flavours'', which are presented on Table
\ref{table:bcast_models}. Most of these variations are clearly expensive,
while others have only an ``historical'' interest. Hence, we chose
for the experiments from Section \ref{sub:Broadcast_practical} two
of the most efficient techniques, the Binomial and the Segmented Chain
Broadcasts, and the simplest one, the Flat Tree Broadcast. 

\begin{table}

\caption{\label{table:bcast_models}Communication models for Broadcast}

\begin{onehalfspace}
\begin{center}\begin{tabular}{|c|c|}
\hline 
\textbf{\scriptsize Strategy}&
\textbf{\scriptsize Communication Model}\tabularnewline
\hline
\hline 
{\scriptsize Flat Tree}&
{\scriptsize $(P-1)\times g(m)+L$}\tabularnewline
\hline 
{\scriptsize Flat Tree Rendezvous}&
{\scriptsize $(P-1)\times g(m)+2\times g(1)+3 \times L$}\tabularnewline
\hline 
{\scriptsize Segmented Flat Tree}&
{\scriptsize $(P-1)\times(g(s)\times k)+L$}\tabularnewline
\hline 
{\scriptsize Chain}&
{\scriptsize $(P-1)\times(g(m)+L)$}\tabularnewline
\hline 
{\scriptsize Chain Rendezvous}&
{\scriptsize $(P-1)\times(g(m)+2\times g(1)+3 \times L)$}\tabularnewline
\hline 
{\scriptsize Seg. Chain (Pipeline)}
&
{\scriptsize $(P-1)\times(g(s)+L)+$}\tabularnewline
&
{\scriptsize $(g(s)\times(k-1))$}\tabularnewline
\hline 
{\scriptsize Binary Tree}&
{\scriptsize $\leq\lceil log_{2}P\rceil\times(2\times g(m)+L)$}\tabularnewline
\hline 
{\scriptsize Binomial Tree}&
{\scriptsize $\lfloor log_{2}P\rfloor\times g(m)+\lceil log_{2}P\rceil\times L$}\tabularnewline
\hline 
{\scriptsize Binomial Tree Rendezvous}&
{\scriptsize $\lfloor log_{2}P\rfloor\times g(m)+$}\tabularnewline
&
{\scriptsize $\lceil log_{2}P\rceil\times(2\times g(1)+3\times L)$}\tabularnewline
\hline 
{\scriptsize Seg. Binomial Tree}&
{\scriptsize $\lfloor log_{2}P\rfloor\times g(s)\times k+\lceil log_{2}P\rceil\times L$}\tabularnewline
\hline
\end{tabular}\end{center}\end{onehalfspace}

\end{table}

\subsection{\label{sub:Broadcast_practical}Practical Results}

\begin{figure}[h]
\begin{center}\includegraphics[%
  width=1.0\columnwidth,
  keepaspectratio]{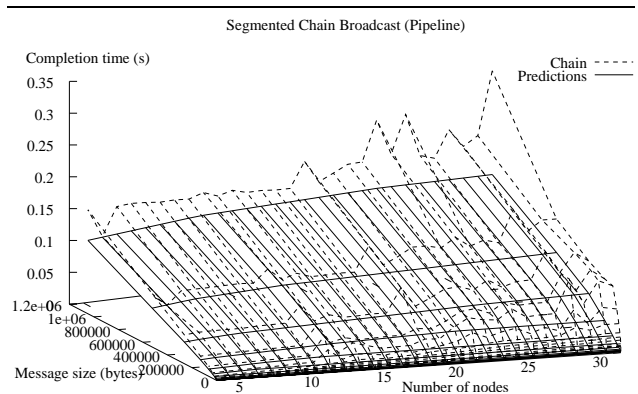}\end{center}
\vspace{-0.3cm}

\caption{\label{Figure:Comparison-Bcast_Chain}Real and expected performance
for the Segmented Chain Broadcast}
\end{figure}

\begin{figure}[h]
\begin{center}\includegraphics[%
  width=1.0\columnwidth,
  keepaspectratio]{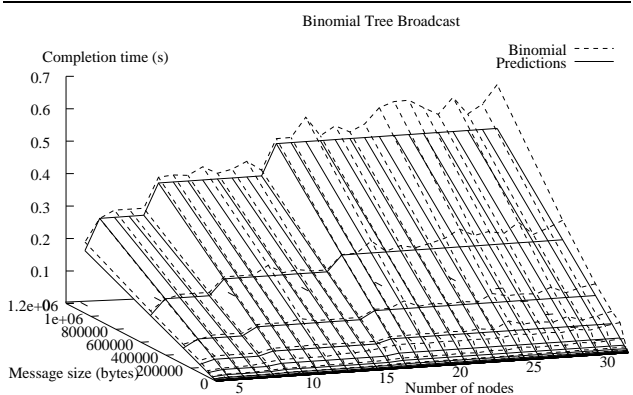}\end{center}
\vspace{-0.3cm}

\caption{\label{Figure:Comparison-Bcast_Bin}Real and expected performance
for the Binomial Broadcast}
\end{figure}

\begin{figure}[h]
\begin{center}\includegraphics[%
  width=1.0\columnwidth,
  keepaspectratio]{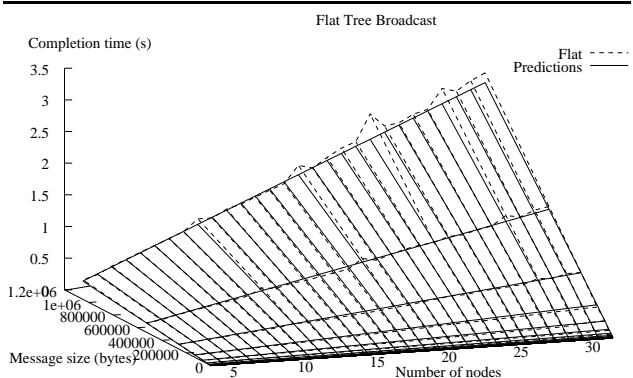}\end{center}
\vspace{-0.3cm}

\caption{\label{Figure:Flat}Real and expected performance for the Flat Tree
Broadcast}
\end{figure}

To evaluate the accuracy of our models, we measured the completion time of the Flat, Binomial
and the Segmented Chain Broadcasts in real experiments, and we compared
these results with the model predictions. Although Flat tree is not
adequate for a large number of processes, we included it because its
simplicity is a good parameter to evaluate other algorithms that use
more complex strategies. Hence, Figures \ref{Figure:Comparison-Bcast_Chain},
\ref{Figure:Comparison-Bcast_Bin} and \ref{Figure:Flat} present
each strategy compared to its performance model's predictions. Despite
some performance variations found mostly in the Segmented Chain and
the Binomial Broadcast, we can observe that predictions seem to follow
the real experiments general behaviour. Actually, as these variations
are much less important in the case of the Flat Broadcast, we think
that they are related to communication delays in some machines, which
are further propagated by the message forwarding, a characteristic
present only on Binomial and Chain broadcasts. As the Flat Tree Broadcast
contacts each node directly, variations in a machine cannot be propagated
to the others, resulting in more accurate predictions, as observed
in Figure \ref{Figure:Flat}. 

\begin{figure}[h]
\begin{center}\includegraphics[%
  width=1.0\columnwidth,
  keepaspectratio]{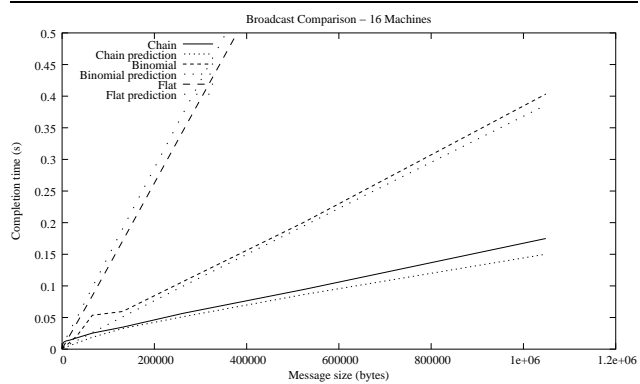}\end{center}
\vspace{-0.3cm}

\caption{\label{Figure:Comparison-between-models Bcast}Comparison between
models and real results, for 16 machines}
\end{figure}

\begin{figure}
\begin{center}\includegraphics[%
  width=1.0\columnwidth]{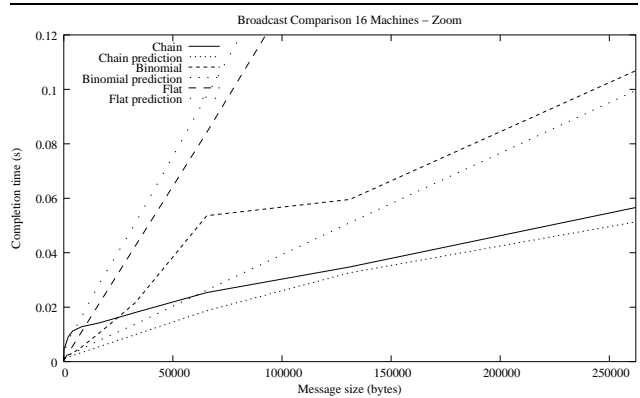}\end{center}
\vspace{-0.3cm}

\caption{\label{Fig: Bcast zoom}Detail on performance degradation with small
messages}
\end{figure}

Figures \ref{Figure:Comparison-Bcast_Chain}, \ref{Figure:Comparison-Bcast_Bin}
and \ref{Figure:Flat}, however, are not in the same scale due to
the different performance level of each algorithm. To compare these
algorithms and to better observe the models' accuracy, we present
on Figure \ref{Figure:Comparison-between-models Bcast} the results
obtained for a group of 16 machines. Here, we observe that the Segmented
Chain Broadcast is the better adapted strategy for our cluster, even
if the models predictions have slightly underestimated the communication
cost. While the observed error rate does not interfere in the selection
process, our attention was drawn by the unexpected delay presented
by the Binomial broadcast when messages are small. A close look on
small messages, as presented in Figure \ref{Fig: Bcast zoom}, shows
that not only the Binomial Broadcast was affected, but also the Segmented
Chain Broadcast. Although this variation does not affect the choice
on the best algorithm, we decided to investigate it closer.

In fact, similar discrepancies were already observed by the LAM-MPI
team \cite{key-14}, and according to Loncaric \cite{key-4}, they
can be due to the TCP acknowledgement policy in some Linux versions.
This problem may delay the transmission of some small messages even
when the TCP\_NODELAY socket option is active (actually, only one
every \emph{n} messages is delayed, with \emph{n} varying from kernel
to kernel). It is true that these effects were mostly present in Linux
kernels 2.0.x and 2.2.x, but according to Loncaric \cite{key-4},
it seems that ``anecdotal evidence suggests that the improved TCP
stack in Linux 2.4 may have problems with many-to-many communication
patterns even though each point-to-point link performs fine''. 

However, if this problem affects the transmission of small messages,
it should also affect the Segmented Chain Broadcast with any message
size, as large messages are split in segments with relatively small
sizes. As the delay observed in Figure \ref{Fig: Bcast zoom} does not seem
to be much evident in the case of Segmented Chain, we believe that
this problem is also related to the management of the send buffers.
We think that the arrival of successive segments forces the transmission
of the messages, masking the undesirable effects when messages are
larger. We plan to answer this question through the investigation
of the segmented variations of the Flat and the Binomial Broadcast,
which similarly to the Segmented Chain, have to deal with small messages
but send many more messages than their traditional versions.

\section{\label{sec:Scatter}Personalised One-to-Many: Scatter}

The Scatter operation, which is also called ``personalised broadcast'',
is an operation where the \emph{root} holds \emph{$m\times P$} data
items that should be equally distributed among the P processes, including
itself. As this is exactly the opposite operation from the Gather primitive, once modelling the Scatter we have a good approximation with the Gather model, which represents the "Many-to-One" communication pattern.

In the case of Scatter, whose root holds a different message for each process, it is believed that optimal algorithms for homogeneous networks use flat trees \cite{key-9}, and by this reason, the Flat Tree approach is the \emph{default} Scatter implementation in most MPI implementations. 

Actually, any other alternative to perform Scatter parallelising the communications requires the transmission of large sets of data to the auxiliary processes, because messages are not identical. Taking for example
the Binomial tree, the root will send down the tree ``bulk'' messages
composed by subsets of the total data. 
Because this strategy allows parallel sends, the completion
time could be reduced, but because the ``bulk'' messages are larger than
a simple message, they take more time to be sent. Hence, the efficiency
of the Binomial Scatter strategy depends on how good the network deals
with large messages, and how the trade-off between parallel sends and 
transmission of large messages will affect the completion time. 

Table \ref{table:Models-for-Scatter} presents the communication model
we constructed for the strategies presented above, and in this paper
we compare Flat Scatter and Binomial Scatter in real experiments.
In a first look, a Binomial Scatter is not as efficient as the Flat
Scatter, because each node receives from the parent node its message
as well as the set of messages it shall send to its successors. On
the other hand, the cost to send these ``combined'' messages (where
most part is useless to the receiver and should be forwarded again)
may be compensated by the possibility to execute parallel transmissions.
As the trade-off between transmission cost and parallel sends is represented
in our models, we can evaluate the advantages of each strategy according
to the clusters' characteristics. 

\begin{table}

\caption{\label{table:Models-for-Scatter}Communication models for Scatter}

\begin{center}\begin{tabular}{|c|c|}
\hline 
\textbf{\scriptsize Strategy}&
\textbf{\scriptsize Communication Model}\tabularnewline
\hline
\hline 
{\scriptsize Flat Tree}&
{\scriptsize $(P-1)\times g(m)+L$}\tabularnewline
\hline 
{\scriptsize Chain}&
{\scriptsize $\sum_{j=1}^{P-1}g(j\times m)+(P-1)\times L$}\tabularnewline
\hline 
{\scriptsize Binomial Tree}&
{\scriptsize $\sum_{j=0}^{\lceil log_{2}P\rceil-1}g(2^{j}\times m)+\lceil log_{2}P\rceil\times L$}\tabularnewline
\hline
\end{tabular}\end{center}
\end{table}

\subsection{\label{sub:Scatter_practical}Practical Results}

\begin{figure}[h]
\begin{center}\includegraphics[%
  width=1.0\columnwidth]{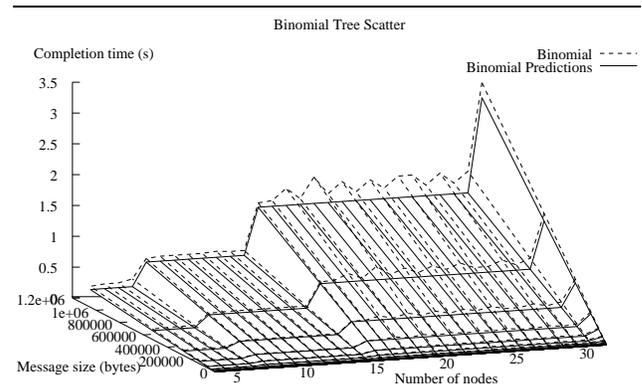}\end{center}
\vspace{-0.3cm}

\caption{\label{Figure:Comparison-Scatter-Bin}Real and expected performance
for the Binomial Scatter}
\end{figure}

\begin{figure}[h]
\begin{center}\includegraphics[%
  width=1.0\columnwidth]{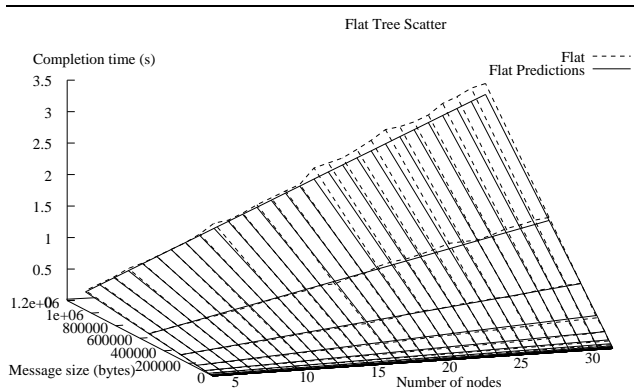}\end{center}
\vspace{-0.3cm}

\caption{\label{Figure:Comparison-Scatter-Flat}Real and expected performance
for the Flat Scatter}
\end{figure}

\begin{figure}[h]
\begin{center}\includegraphics[%
  width=1.0\columnwidth]{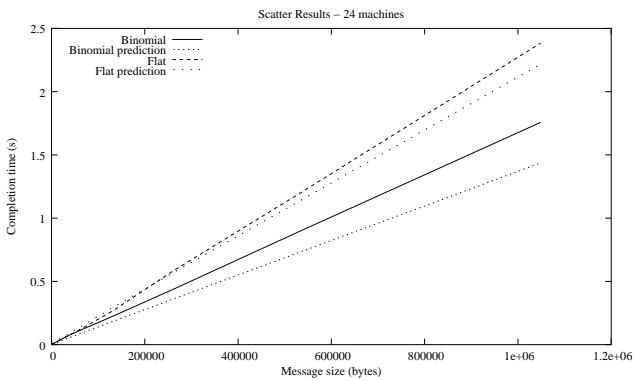}\end{center}
\vspace{-0.3cm}

\caption{\label{Figure:Comparison-between-models-Scatter}Comparison between
Scatter models and real results, for 24 machines}
\end{figure}

In the case of Scatter, we compare the experimental results from Flat
and Binomial Scatters with the predictions from their models. Due to
our network characteristics, our experiments shown that a Binomial
Scatter can be more efficient than Flat Scatter, a fact that is not
usually explored by traditional MPI implementations. As a Binomial
Scatter should balance the cost of combined messages and parallel
sends, it might occur, as in our experiments, that its performance
outweighs the ``simplicity'' from the Flat Scatter with considerable
gains according to the message size and number of nodes, as shown
Figures \ref{Figure:Comparison-Scatter-Bin} and \ref{Figure:Comparison-Scatter-Flat}.
In fact, the Binomial Scatter performance depends on the number of
processes, which gives its characteristic ``stair'' shape, while the
Flat Tree model, limited by the time the root needs to send successive
messages to different nodes (the gap), follows a more linear behaviour.
The varying trade-off on the Binomial Scatter algorithm encourages
the use of our models to identify which implementation is the better
adapted to a specific environment and a set of parameters (message
size, number of nodes), as shown in Figure \ref{Figure:Comparison-between-models-Scatter}.

Nevertheless, Figure \ref{Figure:Comparison-between-models-Scatter}
shows that the models, especially in the case of the Binomial Scatter,
could not avoid a certain level of imprecision. We believe that this
difference is mostly due to the manipulation of large amount of data,
which in the case of the Binomial Scatter is heavily required due
to the ``combined'' messages the nodes receive and forward. 

\vspace{-0.3cm}
\section{\label{sec:Alltoall}Many-to-Many: All to All}

The most intensive and one of the most important communication patterns
for scientific applications is the complete exchange, or All-to-All.
There are several concrete problems whose parallel or distributed
algorithms alternate periods of computing with periods of data exchange
among the processing nodes, with different messages for each other
process. Actually, the All-to-All operation performs a transposition
of data stored across a set of processes, because every process \emph{}holds
\emph{$m\times P$} data items that should be equally distributed
among the P processes, including itself. 

There are many works that focus on the optimisation of All-to-All
and its variant All-to-All-v, where messages can have arbitrary sizes.
Most of these proposals are adapted only to specific network structures,
like meshes, toroids and hypercubes \cite{key-35}. General solutions,
like those found in well known MPI distributions, consider that each
process engages a point-to-point communication with each other, and by consequence, 
the simplest algorithm for All-to-All is called Direct Exchange, where 
all sends and receives are started simultaneously. 

An example of implementation of the Direct Exchange algorithm is the
LAM 6.5.2 MPI\_Alltoall \cite{key-41}. A problem with this algorithm,
however, is that processes usually start communication in the same
order, and consequently, may overload a link by simultaneously sending
messages to a single process each ``round''. Hence, a little optimisation
consists on rotating the communication order
from each process, 
as now implemented in both LAM 7.0.4 \cite{key-32} and MPICH 1.2.5
\cite{key-40}. In spite of this optimisation, that avoids
the overload of a specific process, both strategies do not minimise
communication, and by consequence, communication congestion is highly
probable when the number of nodes increases.

Thus, a major challenge on modelling the communication performance of the All-to-All
operation is the influence of network contention. Models like those presented
by \cite{key-35} are simply extension to the Scatter model that do not take in account the specificities of the All-to-All communication
pattern, nor the non-deterministic behaviour of the network contention. 

Although non-deterministic behaviours are difficult to model, \cite{key-46} introduced a
simple mean to account contention in shared networks, such as non-switched
Ethernet, consisting in a contention factor $\gamma$ that augments
the linear communication model T:

\[T=l+\frac{b\gamma}{W}\]
\vspace{-0.2cm}

where \emph{l} is the link latency, \emph{b} is the message size and
W is the bandwidth of the link, and $\gamma$ is equal to the number
of processes. Using this approach, they found that this simple contention
model greatly enhanced the accuracy of their predictions for essentially
zero extra effort.

Similarly, we assume that contention is sufficiently linear to be modelled. Our approach, however, consists on identifying the performance bounds for the All-to-All
operation, and deriving a relation between such bounds that fits with the 
experimental results for the All-to-All operation. As this ratio depends on
the network characteristics, it is a ``signature'' of such network, and 
therefore can be used in further predictions to obtain results with a considerable precision.

Our performance bounds were also defined as an
extension to the Scatter model, but they considered the main restrictions
to the communication in the all-to-all pattern, specially the nodes' capacity to overlap sends
and receives. Indeed, we explore the fact that even if two messages cannot 
be sent consecutively
in less than \emph{g} through the same link, it takes only \emph{os}
to send a message (more specifically, to deliver the message to the
network card) and \emph{or} to receive it. Consequently, a lower
bound represents the capability to access the network interface as
soon as the precedent send operation returned, while in the upper bound
a node needs to serialise its transmissions due to the link contention. 
These two limits are represented on Table \ref{table:Models-for-Alltoall}.

\begin{table}

\caption{\label{table:Models-for-Alltoall}Communication bounds for the All-to-All operation}

\begin{center}\begin{tabular}{|c|c|}
\hline 
&
\textbf{\scriptsize Communication Model}\tabularnewline
\hline
\hline 
{\scriptsize Upper Bound}&
{\scriptsize $(P-1)\times g(m)+(P-1)\times or(m)+L$}\tabularnewline
\hline 
{\scriptsize Lower Bound}&
{\scriptsize $(P-1)\times os(m)+(P-1)\times or(m)+L$}\tabularnewline
\hline
\end{tabular}\end{center}
\end{table}

\subsection{\label{sub:Alltoall-Practical-Results}Practical Results}

\begin{figure}[h]
\begin{center}\includegraphics[%
  width=1.0\columnwidth]{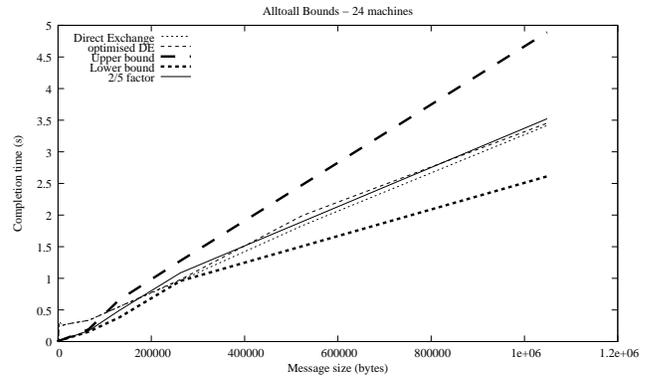}\end{center}
\vspace{-0.3cm}

\caption{\label{Figure:Comparison-bounds}Algorithms performance compared
to All-to-All performance bounds, for 24 machines}
\end{figure}

To illustrate our approach to represent the All-to-All operation in an
environment subjected to network contention, we present, in 
Figure \ref{Figure:Comparison-bounds}, a comparison among the measured 
performance for both Direct Exchange algorithm and its optimised version with
the predicted performance bounds for a group of 24 machines. It can
be observed that both algorithms behave almost identically, and that their
performance differs from the "Scatter-based" model (Lower bound) in a non-negligible amount, which indicates the influence of network contention.

\begin{figure}[h]
\begin{center}\includegraphics[%
  width=1.0\columnwidth]{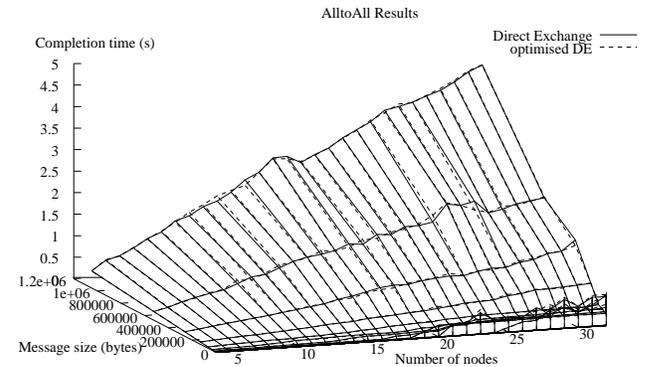}\end{center}
\vspace{-0.3cm}

\caption{\label{Figure:Comparison-Alltoall}Comparison between two All to
All algorithms }
\end{figure}

In fact, the analysis conducted by Grove \cite{key-36} indicated
that ``slow completion times were due to packet losses and their
associated TCP/IP retransmit timeout, caused by extreme network load''. 
Another fact that corroborates Grove's observations is the similarity 
between the Direct Exchange and the Optimised Direct Exchange 
performances (Figure \ref{Figure:Comparison-Alltoall}). This result clearly
indicates that the contention in our experiments comes from the network itself,
and not from the overload of a specific machine.

Therefore, we were able to determine a ratio between the 
predicted Upper and Lower bounds that provides good predictions on the 
performance of the All-to-All operation. This contention ratio $\gamma$ 
is constant and depends only on the network characteristics, whilst the Lower 
and Upper bounds depend on the number of processes, giving a predicted performance of: 

\vspace{-0.2cm}
\[T=Lower+(Upper-Lower) \times \gamma\]
\vspace{-0.2cm}

As a result of our practical experiments, the contention ratio that better 
represents the characteristics of our network was assumed to be 
$\gamma=\frac{2}{5}$. The predicted performances fit with most of the observed 
results, with a small variation only in the case of small messages, which are 
also subjected to the TCP Acknowledgement problem discussed on Section 
\ref{sub:Broadcast_practical}.

This way, despite the non-deterministic behaviour of the network contention, 
we adopted a linear approach where a constant factor, 
characteristic to each network, allows the generation of accurate prediction results.  

\vspace{-0.2cm}

\vspace{-0.2cm}
\section{\label{sec:Conclusions}Conclusions and Future Works}

Existing works that explore the optimisation of heterogeneous networks
usually focus only the optimisation of inter-cluster communication.
We do not agree with this approach, and we suggest to optimise both
inter-cluster and intra-cluster communication. 

For instance, in this paper we propose the use of performance models
to decide, among well known techniques for collective communication,
which is the better adapted for a specific set of parameters (number
of processes, message size). 

As our approach suggests the use of communication models to allow
a fast performance prediction, its accuracy needed to be validated.
Consequently, in this paper we presented three cases that compare
the models' predicted performances and the real results for three
collective communication patterns - ``one-to-all'', ``personalised one-to-all''
and ``many-to-many''. We verified that
the models we construct were accurate enough to predict the performance of the
collective communications, and to allow the selection
of the implementation strategy that better adapts to our network. 

For the modelling of the All-to-All operations, we chose to represent the
effects of network contention as a linear factor. 
Although our experiments demonstrate that linear assumptions were 
accurate enough to predict the performance of such operation, we
agree that this approach does not cover all possibilities in a real
environment. Even though, the results presented in this work offers many 
clues to future investigations on the modelling of communication operations
subjected to non-deterministic network contention behaviours. 

In parallel, we should continue our research on grid-aware collective
communications. We wish to evaluate the accuracy of our models with
other network interconnections, like Myrinet, and we are especially
interested on the automatic organisation of multi-level collective
communications. Hence, our final objective is to integrate both performance
prediction and wide-area communication optimisation in a highly automated 
collective communication library for grid environments.

\end{document}